\documentclass[12pt]{article}
\usepackage{graphicx}
\def\ffrac#1#2{\textstyle{#1\over#2}\displaystyle}
\begin{document}
\pagestyle{myheadings}
\markright{Boundary conformal field theory}

\parindent 0mm
\parskip 6pt

\title{Boundary Conformal Field Theory\footnote{To appear in
{\sl Encyclopedia of Mathematical Physics}, J.-P.~Fran\c coise,
G.~Naber and T.S.~Tsun, eds. (Elsevier, 2005.)}}
\author{John Cardy\\
Rudolf Peierls Centre for Theoretical Physics\\
         1 Keble Road, Oxford OX1 3NP, U.K.\\
and All Souls College, Oxford.}
\maketitle
%
%
\vspace{1cm}

Boundary conformal field theory (BCFT) is simply the study of conformal
field theory (CFT) in domains with a boundary. It gains its significance
because, in some ways, it is mathematically simpler: the
algebraic and geometric structures of CFT appear in a more straightforward
manner; and because it has important applications: in string theory in
the physics of open strings and D-branes, and in condensed matter physics in
boundary critical behavior and quantum impurity models.

In this article, however, I describe the basic ideas from the point of
view of quantum field theory, without regard to particular applications nor
to any deeper mathematical formulations.

\section{Review of CFT}
\subsection{Stress tensor and Ward identities}
Two-dimensional CFTs are massless, local, relativistic
renormalized quantum field theories.
Usually they are considered in imaginary time, i.e. on two-dimensional manifolds
with euclidean signature. In this article the metric is also taken to be
euclidean, although the formulation of CFTs on general Riemann surfaces is
also of great interest, especially for string theory. For the time being
the domain is the entire complex plane.

Heuristically the correlation functions of such a field
theory may be thought of
as being given by the euclidean path integral, that is, as
expectation values of products of
local densities with respect to a Gibbs  measure
$Z^{-1}\,e^{-S_E(\{\psi\})}[d\psi]$, where the $\{\psi(x)\}$
are some set
of fundamental local fields, $S_E$ is the euclidean action, and the
normalization factor $Z$ is the partition function. Of course, such an
object is not in general well-defined, and this picture
should be seen only as a guide to formulating the basic principles of CFT
which can then be developed into a mathematically consistent theory.

In two dimensions, it is useful to use so-called complex coordinates
$z=x^1+ix^2$, $\bar z=x^1-ix^2$.
In CFT there are
local densities $\phi_j(z,\bar z)$, called primary fields,
whose correlation functions transform
covariantly under conformal mappings $z\to z'=f(z)$:
\begin{equation}
\label{cov}
\langle\phi_1(z_1,\bar z_1)\phi_2(z_2,\bar z_2)\ldots\rangle
=\prod_if'(z_j)^{h_j}{\overline f'(z_j)}^{\bar h_j}
\langle\phi_1(z'_1,\bar z'_1)\phi_2(z'_2,\bar z'_2)\ldots\rangle\,,
\end{equation}
where $(h_j,\bar h_j)$ (usually real numbers, not complex conjugates of each
other) are called the \em conformal weights \em of $\phi_j$.
These local fields can in general be normalized so that their two-point
functions have the form
\begin{equation}
\langle\phi_j(z_j,\bar z_j)\phi_k(z_k,\bar z_k)\rangle=
\delta_{jk}/(z_j-z_k)^{2h_j}(\bar z_j-\bar z_k)^{2\bar h_j}\,.
\end{equation}
They satisfy an algebra known as the operator product expansion (OPE)
\begin{equation}
\label{OPE}
\phi_i(z_1,\bar z_1)\cdot\phi_j(z_2,\bar z_2)
=\sum_kc_{ijk}(z_1-z_2)^{-h_i-h_j+h_k}(\bar z_1-\bar z_2)^{-\bar h_i-\bar h_j
+\bar h_k}\phi_k(z_1,\bar z_1)+\cdots\,,
\end{equation}
which is supposed to be valid when inserted into higher-order
correlation functions in the limit when $|z_1-z_2|$ is much less than the
separations of all the other points. The ellipses denote the contributions
of other non-primary scaling
fields to be described below. The structure constants
$c_{ijk}$, along with the conformal weights, characterize the particular CFT.

An essential role is played by the energy-momentum tensor,
or, in euclidean field
theory language, the stress tensor $T^{\mu\nu}$. Heuristically, it is
defined as the response of the partition function to a local change in the
metric:
\begin{equation}
T^{\mu\nu}(x)=-(2\pi)\,\delta\ln Z/\delta g_{\mu\nu}(x)
\end{equation}
(the factor of $2\pi$ is included so that similar factors disappear in later
equations).

The symmetry of the theory
under translations and rotations implies that $T^{\mu\nu}$ is conserved,
$\partial_\mu T^{\mu\nu}=0$, and symmetric. Scale invariance implies that it
is also traceless $\Theta\equiv T^{\mu}_{\mu}=0$. It should be noted that
the vanishing of the trace of the stress tensor for a scale invariant
classical field theory
does not usually survive when quantum corrections are taken into account:
indeed $\Theta\propto\beta(g)$, the renormalization group (RG) beta-function.
A quantum field theory is thus only a CFT when this vanishes, that is at an
RG fixed point.

In complex coordinates the components
$T_{z\bar z}=T_{\bar z z}=4\Theta$ vanish, while the conservation equations
read
\begin{equation}
\partial_{\bar z}T_{zz}=\partial_zT_{\bar z\bar z}=0\,.
\end{equation}
Thus correlators of $T(z)\equiv T_{zz}$ are locally analytic (in fact,
globally
meromorphic) functions of $z$, while those of $\overline T(\bar z)\equiv
T_{\bar z\bar z}$ are anti-analytic. It is this property of analyticity which
makes CFT tractable in two dimensions.

Since an infinitesimal conformal transformation $z\to z+\alpha(z)$ induces
a change in the metric, its effect on a correlation function of primary
fields, given by (\ref{cov}),
may also be expressed through an appropriate integral
involving an insertion of the stress tensor. This leads to the \em conformal
Ward identity:\em
\begin{equation}
\label{WI}
\int_C\langle T(z)\prod_j\phi_j(z_j,\bar z_j)\rangle\,\alpha(z)dz
=\sum_j\left(h_j\alpha'(z_j)+\alpha(z_j)(\partial/\partial z_j)\right)
\langle\prod_j\phi_j(z_j,\bar z_j)\rangle\,,
\end{equation}
where $C$ is a contour encircling all the points $\{z_j\}$. (A similar equation
hold for the insertion of $\overline T$.) Using Cauchy's theorem, this
determines the first few terms in the
OPE of $T$ with any primary density:
\begin{equation}
\label{Tphi}
T(z)\cdot\phi_j(z_j,\bar z_j)={h_j\over(z-z_j)^2}\phi(z_j,\bar z_j)
+{1\over z-z_j}\partial_{z_j}\phi(z_j,\bar z_j)+O(1)\,.
\end{equation}
The other, regular, terms in the OPE generate new scaling fields, which are
not in general primary, called descendants.
One way of defining a density to be primary is by the
condition that the most singular term in its OPE with $T$ is a double pole.

The OPE of $T$ with itself has the form
\begin{equation}
\label{TT}
T(z)\cdot T(z_1)={c/2\over(z-z_1)^4}+{2\over(z-z_1)^2}T(z_1)+\cdots\,.
\end{equation}
The first term is present because $\langle T(z)T(z_1)\rangle$ is non-vanishing,
and must take the form shown, with $c$ being some number (which cannot be
scaled to unity, since the normalization of $T$ is fixed by its definition)
which is a property of the CFT. It is known as the \em conformal anomaly number
\em or the \em central charge\em. This term implies that $T$ is not itself
primary. In fact under a finite conformal transformation $z\to z'=f(z)$
\begin{equation}
\label{Trans}
T(z)\to f'(z)^2T(z')+\ffrac c{12}\{z',z\}\,,
\end{equation}
where $\{z',z\}=\big(f'''f'-\frac32{f''}^2\big)/{f'}^2$ is the
Schwartzian derivative.

\subsection{Virasoro algebra}
As with any quantum field theory, the local fields can be realized as
linear operators acting on a Hilbert space. In ordinary QFT, it is customary
to quantize on a constant time hypersurface. The generator of infinitesimal
time translations is the hamiltonian $\hat H$, which itself is independent
of which time slice is chosen, because of time translational symmetry.
It is also given by the integral over the hypersurface of the time-time
component of the stress tensor. In CFT, because of scale invariance, one
may instead quantize on fixed circle of a given radius. The analog of the
hamiltonian is the dilatation operator $\hat D$,
which generates scale transformations.
Unlike $\hat H$, the spectrum of $\hat D$ is usually discrete, even in an
infinite system.
It may also be expressed as an integral over the radial component of the
stress tensor
\begin{equation}
\hat D={1\over 2\pi}\int_0^{2\pi}r\,\hat T_{rr}\,rd\theta
={1\over 2\pi i}\int_Cz\,\hat T(z)dz-{1\over 2\pi i}\int_C\bar z\,
\hat{\overline T}(\bar z)d\bar z
\equiv \hat L_0+\hat{\overline L}_0\,,
\end{equation}
where, because of analyticity,
$C$ can be any contour encircling the origin.
This suggests that one define other operators
\begin{equation}
\hat L_n\equiv{1\over 2\pi}\int_Cz^{n+1}\hat T(z)dz\,,
\end{equation}
and similarly the $\hat{\overline L}_n$. From the OPE (\ref{TT}) then
follows the Virasoro algebra $\cal V$
\begin{equation}
[\hat L_n,\hat L_m]=(n-m)\hat L_{n+m}+\frac c{12}n(n^2-1)\delta_{n+m,0}\,,
\end{equation}
with an isomorphic algebra $\overline{\cal V}$ generated by the
$\hat{\overline L}_n$.

In radial quantization there is a vacuum state $|0\rangle$. Acting on this
with the operator corresponding to a scaling field gives a
state $|\phi_j\rangle\equiv\hat\phi_j(0,0)|0\rangle$ which is an eigenstate
of $\hat D$: in fact
\begin{equation}
\hat L_0|\phi_j\rangle=h_j|\phi_j\rangle\,,\qquad
\hat{\overline L}_0|\phi_j\rangle=\bar h_j|\phi_j\rangle\,.
\end{equation}
From the OPE (\ref{Tphi}) one sees that $|L_n\phi_j\rangle
\propto \hat L_n|\phi_j\rangle$, and, if $\phi_j$ is primary,
$\hat L_n|\phi_j\rangle=0$ for all $n\geq1$.

The states corresponding to a given primary field,
and those generated by acting on these with all the $\hat L_n$
with $n<0$
an arbitrary number of times, form a highest weight representation
of $\cal V$.
However, this is not necessarily irreducible. There may be
\em null vectors\em, which are linear combinations of states at a given
level which are themselves annihilated by all the $\hat L_n$ with $n>0$.
They exist whenever $h$ takes a value from the Kac table:
\begin{equation}
\label{Kac}
h=h_{r,s}={\big(r(m+1)-sm\big)^2-1\over 4m(m+1)}\,,
\end{equation}
with the central charge parametrized as $c=1-6/\big(m(m+1)\big)$, and
$r$, $s$ are non-negative integers. These null states should be projected out,
giving an irreducible representation ${\cal V}_h$.

The full Hilbert space of the CFT is then
\begin{equation}
\label{decomp}
{\cal H}=\bigoplus_{h,\bar h}\,n_{h,\bar h}\,{\cal V}_h
\otimes{\overline{\cal V}}_{\bar h}\,,
\end{equation}
where the non-negative integers $n_{h,\bar h}$ specify how many distinct
primary fields of weights $(h,\bar h)$ there are in the CFT.

The consistency of the OPE (\ref{OPE}) with the existence of null vectors
leads to the fusion algebra of the CFT. This applies separately to the
holomorphic and antiholomorphic sectors, and determines how many copies
of ${\cal V}_c$ occur in the fusion of ${\cal V}_a$ and ${\cal V}_b$:
\begin{equation}
{\cal V}_a\odot{\cal V}_b = \sum_c N^c_{ab}{\cal V}_c\,,
\end{equation}
where the $N^c_{ab}$ are non-negative integers.

A particularly important subset of all CFTs consists of the
\em minimal models\em. These have rational central charge
$c=1-6(p-q)^2/pq$, in which case the fusion algebra closes with a finite
number of possible values $1\leq r\leq q$, $1\leq s\leq p$ in the Kac
formula (\ref{Kac}). For these models, the fusion algebra
takes the form
\begin{equation}
{\cal V}_{r_1,s_1}\odot{\cal V}_{r_2,s_2}=
{\sum_{r=|r_1-r_2|}^{r_1+r_2-1}}'
{\sum_{s=|s_1-s_2|}^{s_1+s_2-1}}'{\cal V}_{r,s}\,,
\end{equation}
where the prime on the sums indicates that they are to be restricted
to the allowed intervals of $r$ and $s$.

There is an important theorem
which states that the only \em unitary \em
CFTs with $c<1$ are the minimal models with $p/q=(m+1)/m$, where
$m$ is an integer $\geq3$.

\subsection{Modular Invariance}
The fusion algebra limits which values of $(h,\bar h)$ might appear in
a consistent CFT, but not which ones actually occur, i.e. the values of the
$n_{h,\bar h}$. This is answered by the requirement of modular invariance
on the torus.
First consider the theory on an infinitely long cylinder,
of unit circumference. This is related to the (punctured) plane by the
conformal mapping $z\to(1/2\pi)\ln z\equiv t+ix$. The result is
a QFT on the circle $0\leq x<1$, in imaginary time $t$. The generator
of infinitesimal time translations is related to that for dilatations in the
plane:
\begin{equation}
\label{H}
\hat H=2\pi\hat D-\frac{\pi c}6=2\pi\big(\hat L_0+\hat{\overline L}_0\big)
-\frac{\pi c}6\,,
\end{equation}
where the last term comes from the Schwartzian derivative in (\ref{Trans}).
Similarly, the generator of translations in $x$, the total momentum operator,
is $\hat P=2\pi\big(\hat L_0-\hat{\overline L}_0\big)$.

A general torus is, up to a scale transformation, a parallelogram with
vertices $(0,1,\tau,1+\tau)$ in the complex plane, with the opposite edges
identified. We can make this by taking a cylinder of unit circumference
and length ${\rm Im},\tau$, twisting the ends by a relative amount
${\rm Re}\,\tau$,
and sewing them together. This means that the partition function of the CFT
on the torus can be written as
\begin{equation}
Z(\tau,\bar\tau)={\rm Tr}\,e^{-({\rm Im}\,\tau)\hat H+i({\rm Im}\,\tau)\hat P}
={\rm Tr}\,q^{\hat L_0-c/24}\,
{\bar q}^{\hat{\overline L}_0-c/24}\,,
\end{equation}
 using the above expressions for $\hat H$ and $\hat P$ and
introducing $q\equiv e^{2\pi i\tau}$.

Through the decomposition (\ref{decomp}) of $\cal H$,
the trace sum can be written as
\begin{equation}
\label{char}
Z(\tau,\bar\tau)=\sum_{h,\bar h}n_{h,\bar h}\,\chi_h(q)\,\chi_{\bar h}(q)\,,
\end{equation}
where
\begin{equation}
\chi_h(q)\equiv{\rm Tr}_{{\cal V}_h}\,q^{\hat L_0-c/24}
=\sum_Nd_h(N)\,q^{h-(c/24)+N}
\end{equation}
is the character of the representation of highest weight $h$, which counts
the degeneracy $d_h(N)$ at level $N$. It is purely an algebraic property
of the Virasoro algebra, and its explicit form is known in many cases.

\begin{figure}
\centering
\includegraphics[width=8cm]{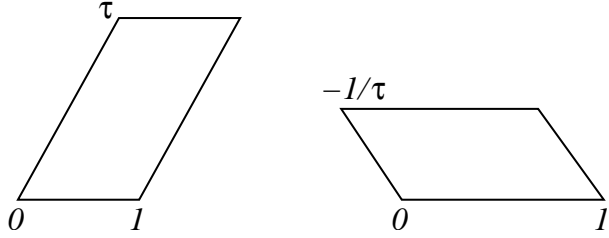}
\caption{\label{figtorus}\small
Two equivalent parametrizations of the same torus.}
\end{figure}

All of this would be less interesting were it not for the observation
that the parametrization of the torus through $\tau$ is not unique. In fact
the transformations $S:\tau\to-1/\tau$ and $T:\tau\to\tau+1$ give the same
torus (see Fig.~\ref{figtorus}).
Together, these operations generate the modular group
SL$(2,{\bf Z})$, and the partition function $Z(\tau,\bar\tau)$ should be
invariant under them. $T$-invariance
is simply implemented by requiring that $h-\bar h$ is an integer, but the
$S$-invariance of the right hand side of (\ref{char}) places highly nontrivial
constraints on the $n_{h,\bar h}$. That this can be satisfied at all relies
on the remarkable property of the characters that they transform linearly
under $S$:
\begin{equation}
\label{mod}
\chi_h(e^{-2\pi i/\tau})=\sum_{h'}S^{h'}_h\chi_{h'}(e^{2\pi i\tau})\,.
\end{equation}
This follows from applying the Poisson sum formula to the explicit
expressions for the characters, which are related to Jacobi theta-functions.
In many cases (for example, the minimal models) this representation is
finite-dimensional, and the matrix $\bf S$ is symmetric and orthogonal.
This means that
one can immediately obtain a modular invariant partition function by
forming the diagonal sum
\begin{equation}
Z=\sum_h\chi_h(q)\chi_h(\bar q)\,,
\end{equation}
so that $n_{h,\bar h}=\delta_{h\bar h}$.
However, because of various symmetries of the characters, other modular
invariants are possible: for the minimal models (and some others) these
have been classified. Because of an analogy of the results with the
classification of semisimple Lie algebras, the diagonal invariants are
called the A-series.

\section{Boundary CFT}
In any field theory in a domain with a boundary, one needs to consider how
to impose a set of consistent boundary conditions. Since CFT is formulated
independently of a particular set of fundamental fields and a lagrangian,
this must be done in a more general manner. A natural requirement is
that the off-diagonal component $T_{\parallel\perp}$ of the stress tensor
parallel/perpendicular to the boundary should vanish. This is called
the conformal boundary condition. If the boundary is
parallel to the time axis, it implies that there is no momentum flow across
the boundary. Moreover, it can be argued that, under the RG, any uniform
boundary condition will flow into a conformally invariant one. For a given
bulk CFT, however, there may be many possible distinct such boundary
conditions, and it is one task of BCFT  to classify these.

To begin with, take the domain to be the upper half plane, so that the
boundary is the real axis. The conformal boundary condition then implies
that $T(z)=\overline T(\bar z)$ when $z$ is on the real axis. This has the
immediate consequence that correlators of $\overline T$ are those of $T$,
analytically continued into the lower half plane. The conformal Ward identity,
\em c.f. \em (\ref{Tphi}), now reads
\begin{eqnarray}
\langle T(z)\prod_j\phi_j(z_j,\bar z_j)\rangle
&=&\sum_j\left({h_j\over(z-z_j)^2}+{1\over z-z_j}\partial_{z_j}\right.
\nonumber\\
&&\qquad
\left.+{\bar h_j\over (\bar z-\bar z_j)^2}+{1\over\bar z-\bar z_j}\partial_{\bar z_j}
\right)\langle\prod_j\phi_j(z_j,\bar z_j)\rangle\,.
\label{BWI}
\end{eqnarray}

In radial quantization, in order that the Hilbert spaces defined on different
hypersurfaces be equivalent, one must choose semicircles centered on some
point on the boundary, conventionally the origin. The dilatation operator
is now
\begin{equation}
\hat D={1\over 2\pi i}\int_Sz\,\hat T(z)dz-{1\over 2\pi i}\int_S\bar z\,
\hat{\overline T}(\bar z)d\bar z\,,
\end{equation}
where $S$ is a semicircle. Using the conformal boundary condition,
this can also be written as
\begin{equation}
\label{DL}
\hat D=\hat L_0={1\over 2\pi i}\int_Cz\,\hat T(z)dz\,,
\end{equation}
where $C$ is a complete circle around the origin. As before, one may similarly
define the $\hat L_n$, and they satisfy a Virasoro algebra.

Note that there is now only one Virasoro algebra. This is related to the
fact that conformal mappings which preserve the real axis correspond to
real analytic functions. The eigenstates of $\hat L_0$ correspond to
\em boundary operators \em $\hat \phi_j(0)$ acting on the
vacuum state $|0\rangle$. It is well-known that in a renormalizable QFT
operators at the boundary require a different renormalization  from those
in the bulk, and this will in general lead to a different set of conformal
weights. It is one of the tasks of BCFT to determine these, for a given
allowed boundary condition.

However, there is one feature unique to boundary CFT in two dimensions.
Radial quantization also makes sense, leading to the same form (\ref{DL})
for the dilation operator, if the boundary conditions on the negative
and positive real axes are different. As far as the structure of BCFT goes,
correlation functions with this mixed boundary condition behave as though
a local scaling field were inserted at the origin. This has led to
the term `boundary condition changing (bcc) operator', but it must be stressed
that these are not local operators in the conventional sense.

\section{The annulus partition function}
\begin{figure}
\centering
\includegraphics[width=5cm]{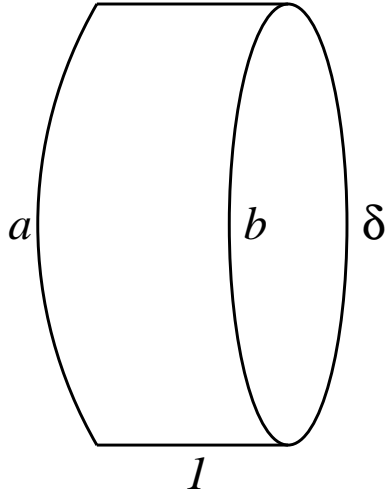}
\caption{\label{figannulus}\small The annulus, with boundary conditions
$a$ and $b$ on either boundary.}
\end{figure}

Just as consideration of the partition function on the torus illuminates
the bulk operator content $n_{h,\bar h}$, it turns out that consistency
on the annulus helps classify both the allowed boundary conditions, and
the boundary operator content. To this end, consider a CFT in an annulus formed
of a rectangle of unit width and height $\delta$, with the top and bottom
edges identified (see Fig.~\ref{figannulus}).
The boundary conditions on the left and right edges, labelled by $a,b,\ldots$,
may be different. The partition function
with boundary conditions $a$ and $b$ on either edge is denoted by
$Z_{ab}(\delta)$.

One way to compute this is by first considering the CFT on an infinitely
long strip of unit width. This is conformally related to the upper half plane
(with an insertion of boundary condition changing operators at $0$ and
$\infty$ if $a\not=b$)
by the mapping $z\to(1/\pi)\ln z$. The generator of infinitesimal translations
along the strip is
\begin{equation}
\hat H_{ab}=\pi\hat D-\pi c/24=\pi\hat L_0-\pi c/24\,.
\end{equation}
Thus for the annulus
\begin{equation}
Z_{ab}(\delta)={\rm Tr}\, e^{-\delta\,\hat H_{ab}}
={\rm Tr}\,q^{\hat L_0-\pi c/24}\,,
\end{equation}
with $q\equiv e^{-\pi\delta}$. As before, this can be decomposed into
characters
\begin{equation}
\label{Zab}
Z_{ab}(\delta)=\sum_hn^{ab}_h\,\chi_h(q)\,,
\end{equation}
but note that now the expression is linear. The non-negative integers
$n_{ab}^h$ give the operator content with the boundary conditions
$(ab)$: the lowest value of $h$ with $n_{ab}^h>0$
gives the conformal weight of the bcc
operator, and the others give conformal weights of the other allowed
primary fields which may also sit at this point.

On the other hand, the annulus partition function may be viewed, up to
an overall rescaling, as the
path integral for a CFT on a circle of unit circumference, being
propagated for (imaginary) time $\delta^{-1}$. From this point of view,
the partition function is no longer a trace, but rather the matrix element
of $e^{-\hat H/\delta}$ between \em boundary states\em:
\begin{equation}
\label{Zab2}
Z_{ab}(\delta)=\langle a|e^{-\hat H/\delta}|b\rangle\,.
\end{equation}
Note that $\hat H$ is the same hamiltonian that appears in (\ref{H}),
and the boundary states lie in $\cal H$, (\ref{decomp}).

How are these boundary states to be characterized? Using the transformation
law (\ref{Trans}) the conformal boundary condition applied to the circle
implies that $L_n=\overline L_{-n}$. This means that any boundary state
$|B\rangle$ lies in the subspace satisfying
\begin{equation}
\label{LL}
\hat L_n|B\rangle=\hat{\overline L}_{-n}|B\rangle\,.
\end{equation}
Moreover, because of the decomposition (\ref{decomp}) of $\cal H$,
$|B\rangle$ is also some linear superposition of states
from ${\cal V}_h\otimes\overline{\cal V}_{\bar h}$.
This condition can therefore be applied
in each subspace.
Taking $n=0$ in (\ref{LL}) constrains $\bar h=h$.
For simplicity, consider only the diagonal CFTs
with $n_{h,\bar h}=\delta_{h,\bar h}$. It can then be shown
that the solution of (\ref{LL}) is unique and
has the following form. The subspace
at level $N$ of ${\cal V}_h$ has dimension $d_h(N)$. Denote an orthonormal
basis by $|h,N;j\rangle$, with $1\leq j\leq d_h(N)$, and the same basis
for $\overline{\cal V}_h$ by $\overline{|h,N;j\rangle}$. The solution
to (\ref{LL}) in this subspace is then
\begin{equation}
|h\rangle\rangle\equiv\sum_{N=0}^\infty\sum_{j=1}^{d_h(N)}
|h,N;j\rangle\otimes\overline{|h,N;j\rangle}\,.
\end{equation}
These are called Ishibashi states. Matrix elements of the translation
operator along the cylinder between them are simple:
\begin{eqnarray}
&&\langle\langle h'|e^{-\hat H/\delta}|h\rangle\rangle\\
&&=\sum_{N'=0}^\infty\sum_{j'=1}^{d_{h'}(N')}
\sum_{N=0}^\infty\sum_{j=1}^{d_h(N)}
\langle h',N';j'|\otimes\overline{\langle h',N';j'|}
e^{-(2\pi/\delta)(\hat L_0+\hat{\overline L}_0-c/12)}\nonumber\\
&&\qquad\qquad\qquad\qquad\qquad\qquad\qquad\qquad\qquad
|h,N;j\rangle\otimes\overline{|h,N;j\rangle}\\
&&=\delta_{h'h}
\sum_{N=0}^\infty\sum_{j=1}^{d_h(N)}e^{-(4\pi/\delta)\big(h+N-(c/24)\big)}
=\delta_{h'h}\,\chi_h(e^{-4\pi/\delta})\,.
\end{eqnarray}
Note that the characters which appear are related to those in
(\ref{Zab}) by the modular transformation $S$.

The physical boundary states satisfying
(\ref{Zab}), sometimes called the Cardy states,
are linear combinations of the Ishibashi states:
\begin{equation}
|a\rangle=\sum_h\langle\langle h|a\rangle\,|h\rangle\rangle\,.
\end{equation}
Equating the two different expressions (\ref{Zab},\ref{Zab2})
for $Z_{ab}$, and using the modular
transformation law (\ref{mod}) and the linear
independence of the characters gives the (equivalent) conditions:
\begin{eqnarray}
n^h_{ab}&=&\sum_{h'}S_{h'}^h\langle a|h'\rangle\rangle
\langle\langle h'|b\rangle\,;\label{C1}\\
\langle a|h'\rangle\rangle \langle\langle h'|b\rangle
&=&\sum_hS_h^{h'}n^h_{ab}\,.\label{C2}
\end{eqnarray}
These are called the Cardy conditions. The requirements that the
right hand side
of (\ref{C1}) should give a non-negative integer, and that the
right hand side of
(\ref{C2}) should factorize in $a$ and $b$, give highly nontrivial constraints
on the allowed boundary states and their operator content.

For the diagonal CFTs considered here (and for the nondiagonal minimal models)
a complete solution is possible.
It can be shown that the
elements $S^h_0$ of $\bf S$ are all non-negative,
so one may choose
$\langle\langle h|\tilde 0\rangle=\big(S_0^h\big)^{1/2}$. This defines a
boundary state
\begin{equation}
|\tilde 0\rangle\equiv\sum_h\big(S_0^h\big)^{1/2}|h\rangle\rangle\,,
\end{equation}
and a corresponding boundary condition
such that $n^h_{00}=\delta_{h0}$. Then, for each $h'\not=0$, one may define
a boundary state
\begin{equation}
\langle\langle h|\tilde{h'}\rangle\equiv S^h_{h'}/\big(S_0^{h}\big)^{1/2}\,.
\end{equation}
From (\ref{C1}), this gives $n^h_{h'0}=\delta_{h'h}$.
For each allowed $h'$ in the torus partition function, there is therefore
a boundary state $|\tilde{h'}\rangle$ satisfying the Cardy conditions.
However, there is a further requirement:
\begin{equation}
\label{verl} n^h_{h'h''}=\sum_\ell{S^h_\ell
S^\ell_{h'}S^\ell_{h''}\over S^\ell_0}
\end{equation}
should be a non-negative integer.
Remarkably, this combination of elements of $\bf S$ occurs in the
Verlinde formula,
which follows from considering consistency of the CFT
on the torus. This states that the right hand side of (\ref{verl})
is equal to the fusion algebra
coefficient $N^h_{h'h''}$. Since these are non-negative integers, the
consistency of the above ansatz for the boundary states is consistent.

We conclude that, at least for the diagonal models, there is a bijection
between the allowed primary fields in the bulk CFT and the allowed conformally
invariant boundary conditions. For the minimal models, with a finite number of
such primary fields, this correspondence has been followed through explicitly.

\subsubsection{Example}
The simplest example is the diagonal
$c=\frac12$ unitary CFT corresponding to $m=3$.
The allowed values of the conformal weights are $h=0,\frac12\frac1{16}$, and
\begin{equation}
{\bf S}=\pmatrix{\ffrac12&\ffrac12&\ffrac1{\sqrt2}\cr
                 \ffrac12&\ffrac12&-\ffrac1{\sqrt2}\cr
                 \ffrac1{\sqrt2}&-\ffrac1{\sqrt2}&0\cr}\,,
\end{equation}
from which one finds the allowed boundary states
\begin{eqnarray}
|\tilde0\rangle&=&\ffrac1{\sqrt2}|0\rangle\rangle
+\ffrac1{\sqrt2}|\ffrac12\rangle\rangle
+\ffrac1{2^{1/4}}|\ffrac1{16}\rangle\rangle\,;\\
|\tilde{\ffrac12}\rangle&=&\ffrac1{\sqrt2}|0\rangle\rangle
+\ffrac1{\sqrt2}|\ffrac12\rangle\rangle
-\ffrac1{2^{1/4}}|\ffrac1{16}\rangle\rangle\,;\\
|\tilde{\ffrac1{16}}\rangle&=&
|0\rangle\rangle-|\ffrac12\rangle\rangle\,.
\end{eqnarray}
The nontrivial part of the fusion algebra of this CFT is
\begin{eqnarray}
{\cal V}_{\frac1{16}}\odot{\cal V}_{\frac1{16}}&=&
{\cal V}_{0}+{\cal V}_{\frac12}\\
{\cal V}_{\frac1{16}}\odot{\cal V}_{\frac12}&=&
{\cal V}_{\frac1{16}}\\
{\cal V}_{\frac12}\odot{\cal V}_{\frac12}&=&
{\cal V}_{0}\,,\\
\end{eqnarray}
from which can be read off the
boundary operator content $n^h_{\tilde h}=1$
and $n^0_{\tilde{\frac1{16}}\tilde{\frac1{16}}}=
n^{\frac12}_{\tilde{\frac1{16}}\tilde{\frac1{16}}}=
n^{\frac12}_{\tilde{\frac1{16}}\tilde{\frac1{16}}}=
n^{\frac1{16}}_{\tilde{\frac12}\tilde{\frac1{16}}}=1$.

The $c=\frac12$ CFT is known to describe the continuum limit of the
critical Ising model, in which spins $s=\pm1$ are localized on the sites
of a regular lattice. The above boundary conditions may be interpreted as
the continuum limit of the lattice boundary conditions $s=1$, free and $s=-1$
respectively. Note there is a symmetry of the fusion rules which means that
one could equally well have inverted the ordering of this correspondence.

\section{Other topics}

\subsection{Boundary entropy}
The partition function on annulus of length $L$ and circumference $\beta$
can be thought of as the quantum statistical
mechanics partition function for a 1d QFT in an interval of length $L$, at
temperature $\beta^{-1}$. It is interesting to consider this in the
thermodynamic limit when $\delta=L/\beta$ is large. In that case, only the
ground state of $\hat H$ contributes in (\ref{Zab2}), giving
\begin{equation}
Z_{ab}(L,\beta)\sim \langle a|0\rangle\langle0|b\rangle\,e^{\pi cL/6\beta}\,,
\end{equation}
from which the free energy $F_{ab}=-\beta^{-1}\ln Z_{ab}$ and the entropy
${\cal S}_{ab}=$\\
$-\beta^2(\partial F_{ab}/\partial\beta)$ can be obtained.
The result is
\begin{equation}
{\cal S}_{ab}=(\pi c/3\beta)L+s_a+s_b+o(1)\,,
\end{equation}
where the first term is the usual extensive contribution. The other
two pieces $s_a\equiv\ln(\langle a|0\rangle)$ and
$s_b\equiv\ln(\langle b|0\rangle)$ may be identified as the \em boundary
entropy \em associated with the corresponding boundary states. A similar
definition may be made in massive QFTs. It has been shown that, analogously
to the statement of Zamoldochikov's $c$-theorem in the bulk,
the boundary entropy is a non-increasing function along
boundary RG flows, and is stationary only for conformal boundary states.

\subsection{Bulk-boundary OPE}
The boundary Ward identity (\ref{BWI}) has the implication that, from the
point of view of the dependence of its correlators on $z_j$ and $\bar z_j$,
a primary field $\phi_j(z_j,\bar z_j)$ may be thought of as the product of
two local fields which are holomorphic functions of $z_j$ and $\bar z_j$
respectively. These will satisfy OPEs as $|z_j-\bar z_j|\to0$, with the
appearance of primary fields on the
right hand side being governed by the fusion rules.
These fields are localized on the real axis: they are the boundary operators.
There is therefore a kind of bulk-boundary OPE:
\begin{equation}
\phi_j(z_j,\bar z_j)=\sum_kd_{jk}({\rm Im}\,z_j)^{-h_j-\bar h_j+h_k}
\phi^b_k({\rm Re}\,z_j)\,,
\end{equation}
where the sum on the right hand side
is in principle over all the boundary fields
consistent with the boundary condition, and the coefficients $d_{jk}$ are
analogous to the OPE coefficients in the bulk. As before, they are
non-vanishing only if allowed by the fusion algebra: a boundary field
of conformal weight $h_k$ is allowed only if $N^{h_k}_{h_j\bar h_j}>0$.

For example, in the $c=\frac12$ CFT, the bulk operator with $h=\bar h=
\frac1{16}$ goes over into the boundary operator with $h=0$, or that
with $h=\frac12$, depending on the boundary condition. The bulk operator
with $h=\bar h=\frac12$, however, can only go over into the identity
boundary operator with $h=0$ (or a descendent thereof.)

The fusion rules also apply to the boundary operators themselves. The
consistency of these with bulk-boundary and bulk-bulk fusion rules, as well
as the modular properties of partition functions,
was examined by Lewellen.

\subsection{Extended algebras}
CFTs may contain other conserved currents apart from the stress tensor,
which generate algebras (Kac-Moody, superconformal, W-algebras) which
extend the Virasoro algebra. In BCFT, in addition to the conformal
boundary condition, it is possible (but not necessary) to impose further
boundary conditions relating the holomorphic and antiholomorphic parts
of the other currents on the boundary.
It is believed that all rational CFTs can
be obtained from Kac-Moody algebras via the coset construction. The
classification of boundary conditions from this point of view is fruitful and
also important for applications, but is beyond the scope of this article.

\subsection{Stochastic Loewner evolution}
In recent years, there has emerged a deep connection between BCFT and
conformally invariant measures on curves in the plane which start at
a boundary of a domain. These arise
naturally in the continuum limit of certain statistical mechanics models.
The measure is constructed dynamically as the curve is extended,
using a sequence of random conformal mappings
called stochastic Loewner evolution (SLE). In CFT, the point where the
curve begins can be viewed as the insertion of a boundary operator. The
requirement that certain quantities should be conserved in mean under
the stochastic process is then equivalent to this operator having a
null state at level two. Many of the standard results of CFT correspond to
an equivalent property of SLE.

\subsection*{Acknowledgments}

This article was written while the author was a member of the Institute
for Advanced Study.
He thanks the School of Mathematics and the School of Natural Sciences
for their hospitality. The work was supported by the
Ellentuck Fund.

\end{document}